# Self-powered programmable van der Waals photodetectors with nonvolatile semi-floating gate

Fan Liu[†], Xi Lin[†], Yuting Yan[†], Xuetao Gan[‡,*], Yingchun Cheng[§,*], Xiaoguang Luo[†,*]

[†] Frontiers Science Center for Flexible Electronics (FSCFE), Shaanxi Institute of Flexible Electronics (SIFE) & Shaanxi Institute of Biomedical Materials and Engineering (SIBME), Northwestern Polytechnical University, Xi'an 710129, China

[‡] Key Laboratory of Light Field Manipulation and Information Acquisition, Ministry of Industry and Information Technology, and Shaanxi Key Laboratory of Optical Information Technology, School of Physical Science and Technology, Northwestern Polytechnical University, Xi'an 710129, China

[§] Key Laboratory of Flexible Electronics & Institute of Advanced Materials, Jiangsu National Synergetic Innovation Center for Advanced Materials, Nanjing Tech University, Nanjing 211816, China

[*] Corresponding Authors

xuetaogan@nwpu.edu.cn

iamyccheng@njtech.edu.cn

iamxgluo@nwpu.edu.cn

**Abstract**

Tunable photovoltaic photodetectors are of significant relevance in the fields of programmable and neuromorphic optoelectronics. However, their widespread adoption is hindered by intricate architectural design and energy consumption challenges. This study employs a nonvolatile $MoTe_2$/hBN/graphene semi-floating photodetector to address these issues. Programed with pulsed gate voltage, the $MoTe_2$ channel can be reconfigured from an $n^+$-n to a p-n homojunction, and the photocurrent transition changes from negative to positive values. Scanning photocurrent mapping reveals that the negative and positive photocurrents are attributed to Schottky junction and p-n homojunction, respectively. In the p-n configuration, the device demonstrates self-driven, linear, rapid response (~3 ms), and broadband sensitivity (from 405 to 1500 nm) for photodetection, with typical performances of responsivity at ~0.5 A/W and detectivity ~$1.6\times10^{12}$ Jones under 635 nm illumination. These outstanding photodetection capabilities emphasize the potential of the semi-floating photodetector as a pioneering approach for advancing logical and nonvolatile optoelectronics.

**Keywords:** Photovoltaic effect, tunable, two-dimensional photodetectors, semi-floating gate, nonvolatile

Recently, there has been a growing interest in tunable photovoltaic photodetectors (PDs) utilizing atomically thin two-dimensional (2D) materials. This surge in attention is particularly evident with the advancement of programmable[1, 2] and neuromorphic[3-5] optoelectronics. The photovoltaic response is characterized by attributes such as low-power consumption, rapid response, and low dark current.[6-8] This behavior is commonly attributed to the built-in electric field within various junctions, including $p^+$-p, p-n, $n^+$-n, and Schottky junctions, which are typically fabricated through processes such as chemical doping,[9, 10] materials stacking,[11, 12] epitaxial growth,[13, 14] local split gating,[15-19] strain engineering,[20, 21] or asymmetric contacting.[22-25] Distinguished by the volatile or nonvolatile nature of these junctions, two strategies involving continuous and pulsed voltage application are typically employed to tune the photovoltaic photoresponse, enabling transitions from negative to positive responses. For example, in certain narrow-bandgap 2D semiconductor-based heterojunctions (e.g., BP/$MoTe_2$,[26] $PdSe_2$/$MoTe_2$,[27] AsP/$MoS_2$[28]) and the configuration of ambipolar 2D semiconductor-based homojunctions (e.g., BP,[19] $WSe_2$,[15] $MoTe_2$[29]), the band alignment can be effectively modulated through continuous global gating and split gating, respectively, resulting in gate-tunable photoresponse. However, it is worth noting that continuous gating consumes power. To address this, alternative nonvolatile structures such as ferroelectric polymers,[1, 30, 31] sulfur-vacancy adjustable $MoS_2$,[5] and semi-floating gate (SFG)[32, 33] have been introduced. Among these, the SFG structure stands out due to its well-defined dimensions, prolonged retention time, rapid response, and cost-efficiency.[33, 34] Nonetheless, a comprehensive investigation into the photodetection capabilities of semi-floating gate photodetectors (SFG-PDs) with 2D materials is yet to be conducted, especially in terms of understanding the underlying mechanisms of their photoresponse.

In this study, we present a $MoTe_2$/hBN/graphene semi-floating gate photodetector (SFG-PD) with Au electrodes, demonstrating its capability to modulate photocurrent from negative to positive values through pulsed gate voltage ($V_{\text{cg-pulse}}$). The choice of $MoTe_2$ is based on its ambipolar property and narrow bandgap (~ 0.9 eV for multilayer[35, 36]). Through comprehensive electrical and photoelectrical characterizations, we identified the p-n and $n^+$-n configurations of the $MoTe_2$ channel for positive and negative $V_{\text{cg-pulse}}$, respectively. Specifically, positive photocurrent arises from the p-n homojunction for $V_{\text{cg-pulse}} > 0$, while negative photocurrent stems from the photoresponse at two back-to-back Au/$MoTe_2$ Schottky junctions, rather than at the $n^+$-n homojunction, for $V_{\text{cg-pulse}} < 0$. The SFG-PD in p-n configuration exhibits linear, rapid response (~3 ms), and broadband sensitivity (spanning from 405 to 1500 nm), with a high responsivity of ~0.5 A/W and detectivity ~$1.6\times10^{12}$ Jones at 635 nm. Furthermore, based on fitting results, the photocurrent is projected to decrease to 67.07% after 10 years. These exceptional properties, combined with the ability for tunable photoresponse, position the SFG-PD as an extremely promising candidate for applications in photovoltaics, nonvolatile circuits, and logical systems.

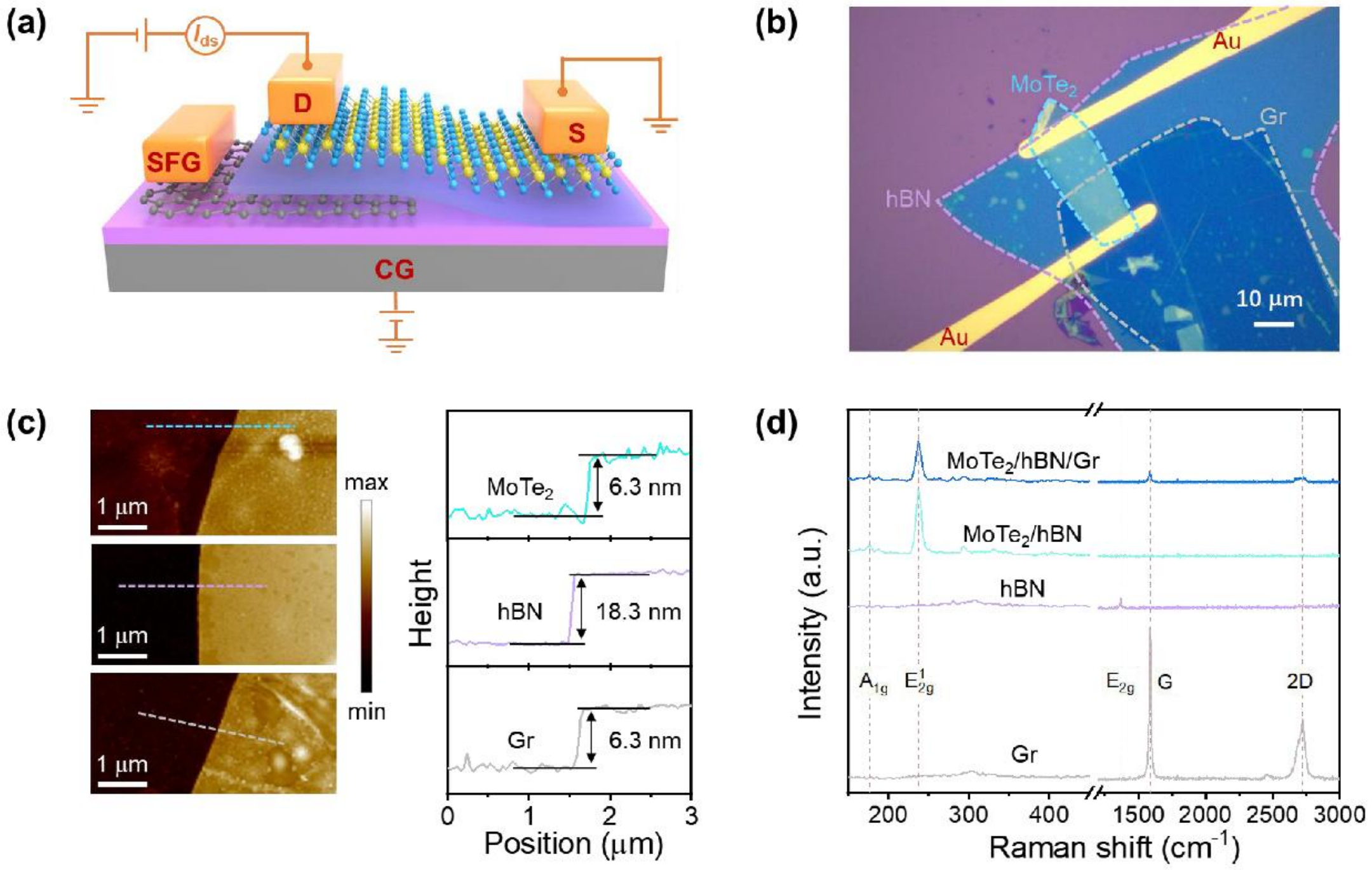

Fig. 1. (a) Schematic of the $MoTe_2$/hBN/Gr-based SFG-PD on a $SiO_2$/Si substrate. (b) Optical microscopic image of a typical device, where the area of conducting channel is about 325 $\mu m^2$. (c) AFM image (left panel) and height profile (right panel) along corresponding lines of $MoTe_2$, hBN, and Gr, respectively. (d) Raman spectra at different position of the $MoTe_2$/hBN/Gr heterostructure.

Mechanically exfoliated 2H-$MoTe_2$, hexagonal boron nitride (hBN), and graphene (Gr) are utilized in the fabrication of the device on a $SiO_2$/Si substrate. A schematic diagram is presented in Fig. 1a. Gr serves as the SFG and is partially covered by the top half of the $MoTe_2$ channel. Simultaneously, hBN functions as the dielectric layer. Detailed information about the fabrication process can be found in the Experimental Section of the Supporting Information. An optical image of a typical $MoTe_2$/hBN/Gr SFG-PD is displayed in Fig. 1b, with $MoTe_2$, hBN, and Gr highlighted by cyan, purple, and gray dashed lines, respectively. The thicknesses of the $MoTe_2$, hBN, and Gr layers, as determined through atomic force microscopy measurements, are 6.3, 18.3, and 6.3 nm, respectively, as shown in Fig. 1c. The presence of these constituent materials and their crystal quality has been confirmed by distinct characteristic peaks in the Raman spectra, as illustrated in Fig. 1d. Specifically, Raman peaks around 1368 $cm^{-1}$ correspond to $E_{2g}$ mode of hBN.[37] The characteristic peaks of Gr are observed in the high-frequency region at around 1585 (G band) and 2723 $cm^{-1}$ (2D band).[38] Additionally, the out-of-plane $A_{1g}$ and in-plane $E_{2g}^{1}$ vibration modes of $MoTe_2$ are detected at 179 and 238 $cm^{-1}$, respectively.[39]

Electrical characterizations were performed using a Si substrate as the control gate ($V_{cg}$) for channel programming and erasure, with Gr serving as the SFG for charge storage. Sweeping of $V_{cg}$ from −40 to 40 V resulted in the appearance of cross hysteresis in transfer curves ($I_{ds} - V_{cg}$) with the double memory windows, as shown in Fig. 2a, indicating the nonvolatile property of the device with double memory windows. Subsequent application of pulsed gate voltage with $V_{cg-pulse} = \pm 40$ (±60) V produced opposite rectification characteristics at positive and negative $V_{cg-pulse}$, as illustrated in output curves ($I_{ds} - V_{ds}$; Fig. 2b). Noteworthy, charge modulation within the hBN/Gr van der Waals floating gate can be achieved with a speed of 20 ns through modified Fowler–Nordheim tunneling.[40] In this study, the pulse duration was set to 1 second to ensure saturation at specific $V_{cg-pulse}$ levels. The rectification ratio is approximately $10^2$ for both $V_{cg-pulse} = -40$ and −60 V, while at 60 V it is about an order of magnitude higher (~$10^4$) than at 40 V (~$10^3$), which indicates different modulation degrees or distinct modulation mechanisms. This effect, observed in the $WSe_2$ SFG field-effect transistor, has been attributed to charge injection/trapping in the SFG and the channel homojunction configuration (p-n and $n^+$-n).[33] However, with Au electrodes in our device, the $n^+$-n homojunction contribution is not predominant. As shown in Fig. S1b, the memory windows widen as the scanning voltage range increases, suggesting that the threshold voltage of the memory characteristic can be adjusted by $V_{cg-pulse}$.[40] This adjustment is also evident in Kelvin probe force microscope (KPFM) measurements. In Fig. S1c–d, the surface potentials above Gr (the band information will not be affected by the hBN layer[41]) at $V_{cg-pulse} = 60$ and −60 V exhibit marked differences, suggesting distinct charge accumulation in the Gr SFG. Importantly, a clear Fermi-level difference (~0.5 eV) at the $MoTe_2$ homojunction was observed at 60 V but not at −60 V, as demonstrated in Fig. S1e–f. Therefore, it is evident that a p-n homojunction is formed at $V_{\text{cg−pulse}} = 60$ V, while the n⁺-n homojunction is inconspicuous at $V_{\text{cg−pulse}} =− 60$ V.

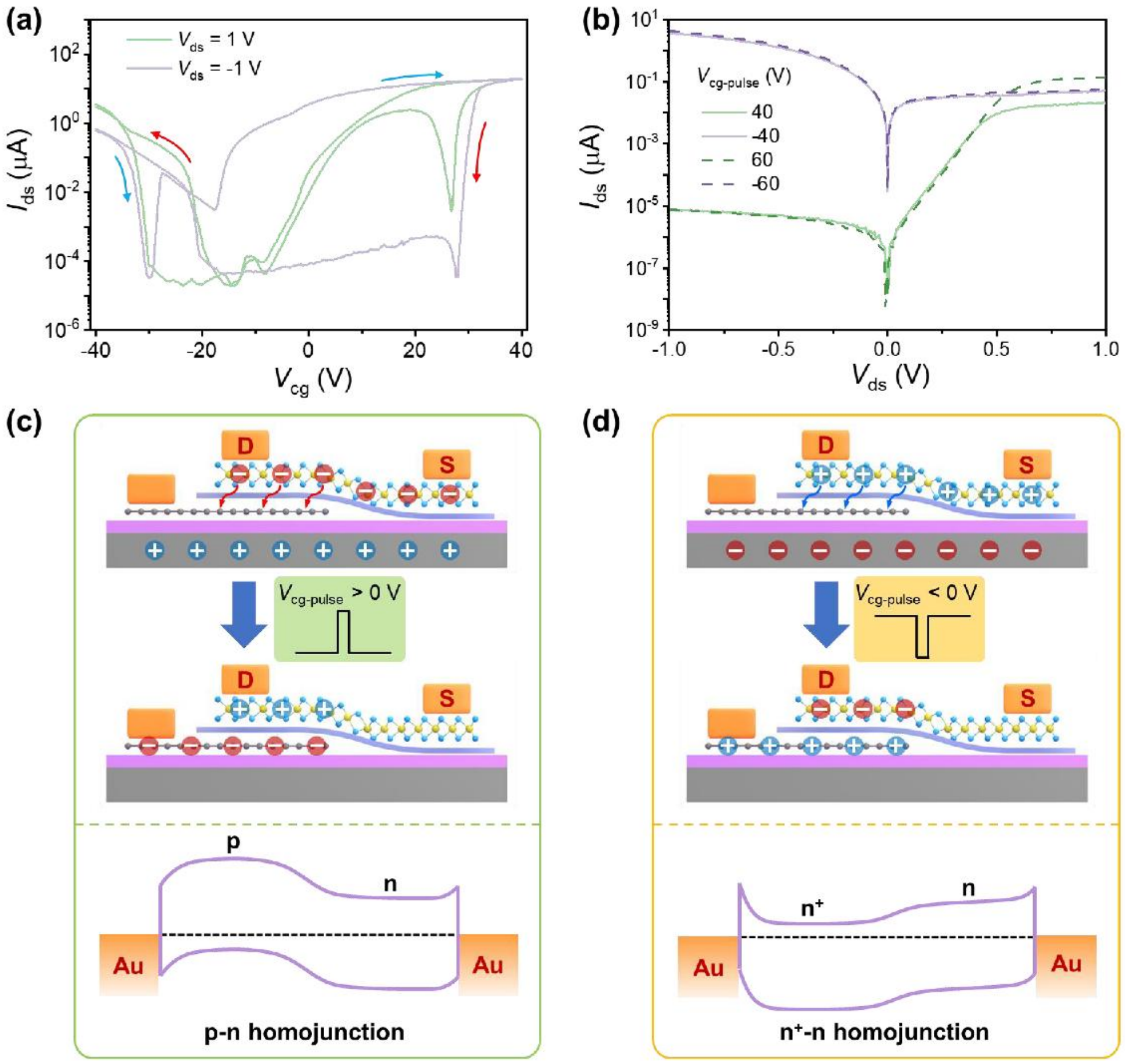

Fig. 2. (a) Semi-log plot of transfer curves ($I_{\text{ds}} - V_{\text{cg}}$) with forward and reverse $V_{\text{cg}}$ scan directions. (b) Semi-log plot of output curves ($I_{\text{ds}} - V_{\text{ds}}$) under $V_{\text{cg}}$ pulse with 60 and −60 V, where pulse duration is 1 s. (c) Operation diagrams of the SFG-PD with positive $V_{\text{cg}}$ applied and released to the Si control gate (top panel), and band diagram of the device in p-n homojunction configuration (bottom panel). (d) Operation diagrams of the SFG-PD with negative $V_{\text{cg}}$ applied and released to the Si control gate (top panel), and band diagram of the device in n⁺-n homojunction configuration (bottom panel).

Schematic diagrams and band structures are employed to explain the opposite rectification at opposite $V_{\text{cg−pulse}}$ values. For $V_{\text{cg−pulse}} > 0$ V, as depicted in Fig. 2c, electrons in $MoTe_2$ tunnel through the hBN layer to Gr with the assistance of the applied electric field, which are then stored in Gr for the SFG. Thus, the Fermi level of the semi-channel above the SFG is lowered, allowing the device to operate as a p-n diode, with the band structure sketched in the bottom panel of Fig.

2c. Likewise, for $V_{\mathrm{cg-pulse}} < 0$ V, holes tunnel toward Gr with the assistance of the applied electric field, and are subsequently stored in Gr for the SFG, resulting in the Fermi-level increase in the semi-channel over the SFG (Fig. 2d). An $n^+$-n homojunction is expected to form in the $MoTe_2$ channel, though it appears to be very weak according to KPFM results. We infer that two back-to-back asymmetric Au/$MoTe_2$ Schottky junctions are critical in the rectification characteristic when $V_{\mathrm{cg-pulse}} =- 60$ V (with the band structure shown in the bottom panel of Fig. 2d), which will be further corroborated by subsequent scanning photocurrent imaging.

The current through a diode with series resistance $R_\mathrm{s}$ can be modeled by the modified Shockley equation[42]

$$I_{\mathrm{ds}} = \frac{nV_\mathrm{T}}{R_\mathrm{s}} W\left[\frac{I_0 R_\mathrm{s}}{nV_\mathrm{T}} \exp\left(\frac{V_{\mathrm{ds}} + I_0 R_\mathrm{s}}{nV_\mathrm{T}}\right)\right] - I_0$$

where $V_\mathrm{T} = k_\mathrm{B}T/q$ is the thermal voltage at temperature $T$, and $W$ is the Lambert $W$ function. $k_\mathrm{B}$, $q$, $I_0$, and $n$ are the Boltzmann constant, electron charge, reverse saturation current, and ideality factor, respectively. By fitting the data shown in Fig. 2b in the range of $0 \leq |V_{\mathrm{ds}}| \leq 0.4$ V, we found $n \sim 1.00$ for both $V_{\mathrm{cg-pulse}} = 40$ and 60 V, and $n \sim 1.44$ (1.34) for $V_{\mathrm{cg-pulse}} =- 40$ (−60) V. Therefore, the carrier transport in the device is primarily governed by a diffusion process for p–n homojunction configuration while being influenced by both the diffusion process and the recombination process for the configuration with $n^+$-n homojunction and Schottky junctions.

To explore the adjustable photoresponse in photovoltaic mode (i.e., without a bias voltage $V_{\mathrm{ds}}$), the SFG-PD was exposed to global illumination by lasers of various wavelengths and intensities. The photogenerated electron-hole pairs in the $MoTe_2$ channel are separated by the built-in electric field of the p-n, $n^+$-n, or Schottky junctions, resulting in a short-circuit current ($I_{\mathrm{sc}}$) and open-circuit voltage ($V_{\mathrm{oc}}$) in the asymmetric electrical environment of the entire conducting channel[24]. Fig. 3a presents output curves of the device under varying laser power densities ($P_{\mathrm{in}}$) when $V_{\mathrm{cg-pulse}} = 60$ V, using a 532 nm wavelength laser with the spot diameter of ~3 mm. With increasing $P_{\mathrm{in}}$, both $I_{\mathrm{sc}}$ and $V_{\mathrm{oc}}$ gradually deviate from zero, indicating a significant photovoltaic effect. At $P_{\mathrm{in}} = 12.68$ mW/cm$^2$, a notable and unsaturated open-circuit voltage of $V_{\mathrm{oc}} = 0.4$ V was reached along with a short-circuit current of $I_{\mathrm{sc}} = 5.48$ nA. The extracted $I_{\mathrm{sc}}$ and $V_{\mathrm{oc}}$ as functions of $P_{\mathrm{in}}$ are presented in Fig. 3b. By fitting the data using the power relationship $I_{\mathrm{sc}} \sim P_{\mathrm{in}}^{\gamma}$, we obtained $\gamma \sim 0.99$, suggesting a strong linear dependence of $I_{\mathrm{sc}}$ on $P_{\mathrm{in}}$.

In an ideal photodiode, the photocurrent is absolutely linear with light intensity, that is, $\gamma = 1.00$. The slight deviation from this strict linear response in our device may be due to disorders, defects, or impurities, as seen in other 2D photodetectors[16, 43]. The ideality factor of $n \sim 1.00$ and the power coefficient of $\gamma \sim 0.99$ confirm the high quality of the $MoTe_2$ p-n homojunction regarding both electric transport and photoresponse. The electrical power ($P_{\mathrm{el}} = I_{\mathrm{ds}} \cdot V_{\mathrm{ds}}$) values are shown in Fig. 3c, with the maximum $P_{\mathrm{el}}$ being ~1.10 nW at $V_{\mathrm{ds}} = 0.26$ V and $P_{\mathrm{in}} = 12.68$ mW/cm$^2$. The maximum $P_{\mathrm{el}}$ values at different $P_{\mathrm{in}}$ are typically used to calculate the power conversion efficiency ($\eta_{\mathrm{PCE}} = P_{\mathrm{el}}/P_{\mathrm{in}}A$), where $A$ represents the active area of the photodetector. In this study, we consider the entire conducting channel as the active area, which means the maximum $\eta_{\mathrm{PCE}} \sim 3.73\%$ obtained at $P_{\mathrm{in}} = 0.25$ mW/cm$^2$ is likely an underestimation, as the actual area of the p-n homojunction is significantly smaller than that of the entire conducting channel. Moreover, we can determine the fill factor ($FF$) using the formula of $FF = P_{\mathrm{el}}/V_{\mathrm{oc}}I_{\mathrm{sc}}$, which varies from approximately 0.27 to 0.65 across the laser power densities, as shown in Fig. 3d.

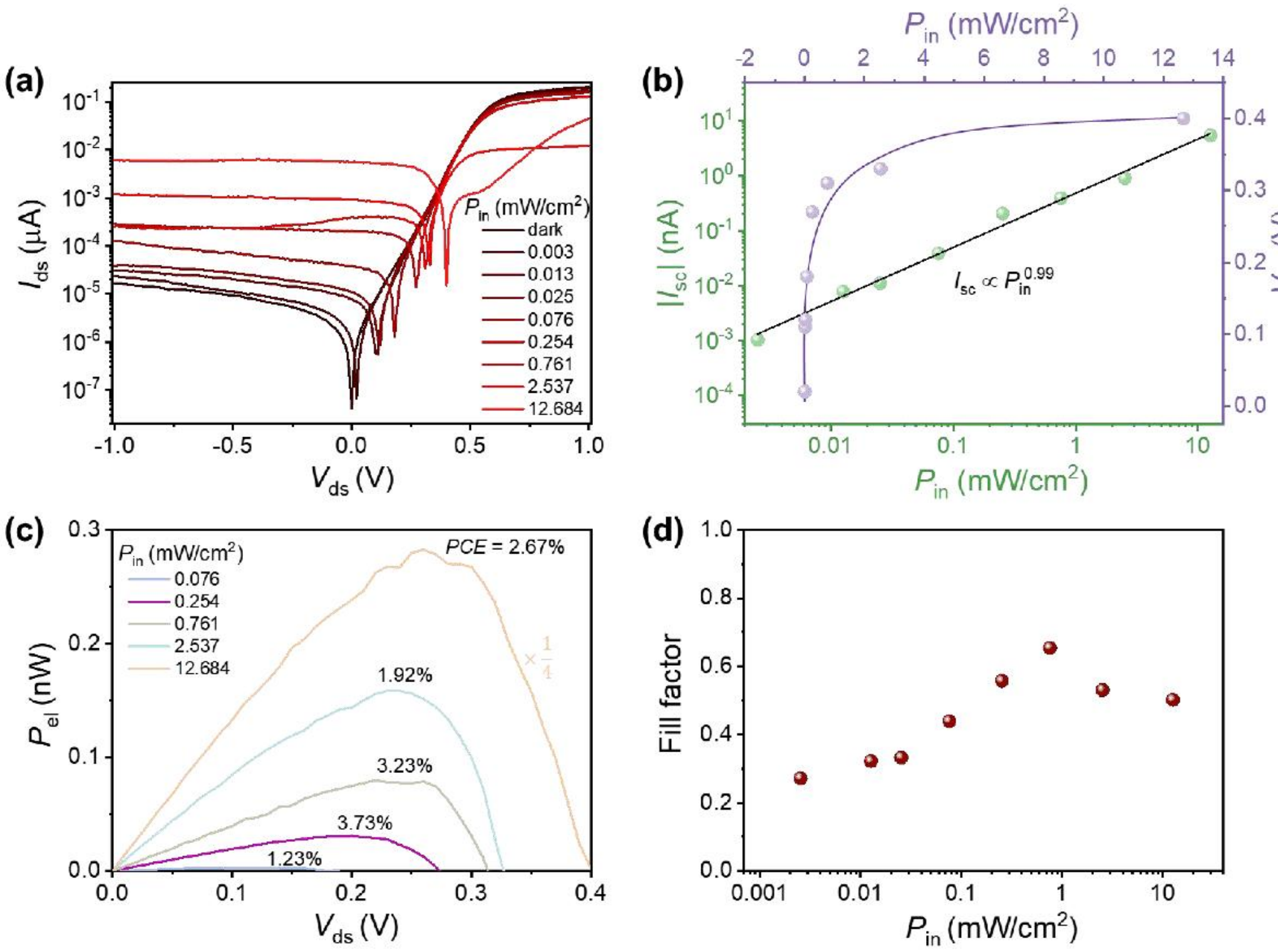

Fig. 3. (a) Semi-log plot of output curves under 532 nm illumination with different $P_{\mathrm{in}}$. (b) Extracted $I_{\mathrm{sc}}$ and $V_{\mathrm{oc}}$ as a function of $P_{\mathrm{in}}$. The calculated electrical power (c) and fill factor (d) at different $P_{\mathrm{in}}$.

With a bandgap ~0.9 eV, the multilayer 2H-$MoTe_2$-based SFG-PD exhibits photoresponse across a broad spectrum ranging from ultraviolet to near-infrared, as corroborated by the output characteristics under illumination with varying laser wavelengths spanning from 405 to 1500 nm (Figs. 3 and S2–S3). Notably, a pronounced photovoltaic effect is observed at laser wavelengths of 405, 532, 635, 785, 1064, 1260, 1350, and 1400 nm. However, a conspicuous decline is noted in both $I_{sc}$ and $V_{oc}$ from 1260 to 1400 nm. When illuminated with a 1500 nm laser, the SFG-PD exhibits virtually no discernible photovoltaic effect. The observed photoresponse behavior aligns well with the limitation imposed by the $MoTe_2$ bandgap and is consistent with the previously reported absorption spectrum of few-layer $MoTe_2$,[35] as introduced in Fig. S3c–d.

To investigate the transient photoresponse, the SFG-PD is globally illuminated using a switchable laser with a modulation frequency of 1 Hz (chosen based on the photoresponse time of the SFG-PD). Figs. 4a and S4 present the time-dependent photoresponse at different laser wavelengths (including 405, 532, 635, 785, and 1064 nm) and laser power densities when $V_{cg-pulse} = 60$ V. Both the photocurrent ($I_{ph}$) and photovoltaic voltage exhibit clear on-off behavior, with the current on-off-ratio reaching up to $10^5$ under strong illumination. It is worth noting, however, that the photoresponse becomes unstable when $P_{in} > 21.06$ mW/cm$^2$ under 405 nm illumination, as indicated in Fig. S4a. This phenomenon can be attributed to photoinduced doping (stemming from light absorption by defects in hBN), resulting in a polarity change in $MoTe_2$ and a consequent weakening of the p-n junction.[44-46] Therefore, the photoresponse under 405 nm illumination when $P_{in} > 21.06$ mW/cm$^2$ is not considered in the subsequent photodetection analysis. Because of the strong built-in electric field at the p-n homojunction, the response time is on the order of milliseconds for all tested wavelengths. Fig. 4b illustrates a typical on-off process under 532 nm illumination, and the rise/decay time ~3.1/2.3 ms, estimated between 10% and 90% of the maximum photocurrent, is closely corresponding to the measured 3dB bandwidth of ~210 Hz (Fig. S6). It is worth noting that the response time can be further reduced by minimizing the $MoTe_2$ channel and optimizing the electrode/$MoTe_2$ contacts.

The photocurrent and photovoltaic voltage extracted from Fig. S4 are depicted in Figs. 4c and 4d, respectively. The photocurrent exhibits a linear dependence on the laser power density (i.e., $\gamma{\sim}1.0$ after fitting by $I_{ph}{\sim}P_{in}^{\gamma}$), which can be attributed to the high quality of the p-n homojunction when $V_{cg-pulse} = 60$ V. However, for the configuration with $n^+$-n homojunction and Schottky junctions at $V_{\mathrm{cg-pulse}} =-60$ V, the linearity may be disrupted by interactions among carriers, photocarriers, and trap states. Fig. S7 presents the photoresponse of a device under 532 nm illumination at positive and negative $V_{\mathrm{cg-pulse}}$, from which one can observe that $\gamma = 1.01$ at $V_{\mathrm{cg-pulse}} = 60$ V while $\gamma = 0.80$ at $V_{\mathrm{cg-pulse}} =-60$ V. Furthermore, the response time for $V_{\mathrm{cg-pulse}} =-60$ V is longer than that at 60 V, indicating lower carrier transition efficiency in the configuration with $n^+$-n homojunction and Schottky junctions.

The photovoltaic voltage can be utilized to investigate the carrier recombination process within the device.[47] As indicated by the formula,

$$\frac{dV_{\mathrm{oc}}}{d\ln(P_{\mathrm{in}})} = \frac{2}{\beta}\frac{k_{\mathrm{B}}T}{q}$$

$\beta$ tends to 1 when $P_{\mathrm{in}} < 1$ mW/cm$^2$ and increases to 2 when $P_{\mathrm{in}} > 1$ mW/cm$^2$ (Fig. 4d), suggesting Shockley-Read-Hall (trap-assisted) recombination predominates under weak illumination, while Langevin (band-to-band) recombination prevails under strong illumination. This transition can be attributed to the indirect bandgap of multilayer $MoTe_2$. In fact, multilayer $MoTe_2$ can exhibit both indirect and direct transitions, though the direct transition is considerably weaker.[36] The prevalence of indirect transitions dictates recombination under weak illumination due to the low density of photogenerated carriers, favoring the trap-assisted process to conserve momentum during electron transitions. As laser power increases, the photogenerated carrier density increases, with more electrons or holes occupying higher energy levels, resulting in a growing contribution from direct transitions.

For a photodetector, responsivity ($R$) and detectivity ($D^*$) are two critical parameters for evaluating detection performance. $R$ characterizes the input-output efficiency of the photodetector and is defined as the ratio of photocurrent to effective light power:

$$R = \frac{I_{\mathrm{ph}}}{P_{\mathrm{in}}A}$$

Meanwhile, $D^*$ can be estimated using the formula:

$$D^* = R\sqrt{\frac{A}{2qI_{\mathrm{dark}}}}$$

assuming that shot noise predominates.[48] After calculation using the data presented in Fig. 4c, we observed that both $R$ and $D^*$ exhibit a near-independence on laser power density, as shown in Fig. 4e. Fig. 4f provides the average values of these parameters concerning laser wavelength, with maximum values approximated at 0.5 A/W and $1.6\times10^{12}$ Jones, respectively, under 635 nm illumination. It is worth noting that these performance parameters may be potentially underestimated due to the overestimated active area of the SFG-PD. Additionally, $R$ and $D^*$ align closely with the calculated absorbance of 6.3 nm $MoTe_2$ (Fig. S8) using the Lambert–Beer law and the reported experimental data of trilayer $MoTe_2$,[36] which can potentially be further enhanced through optical or plasmonic resonances.[21, 49, 50]

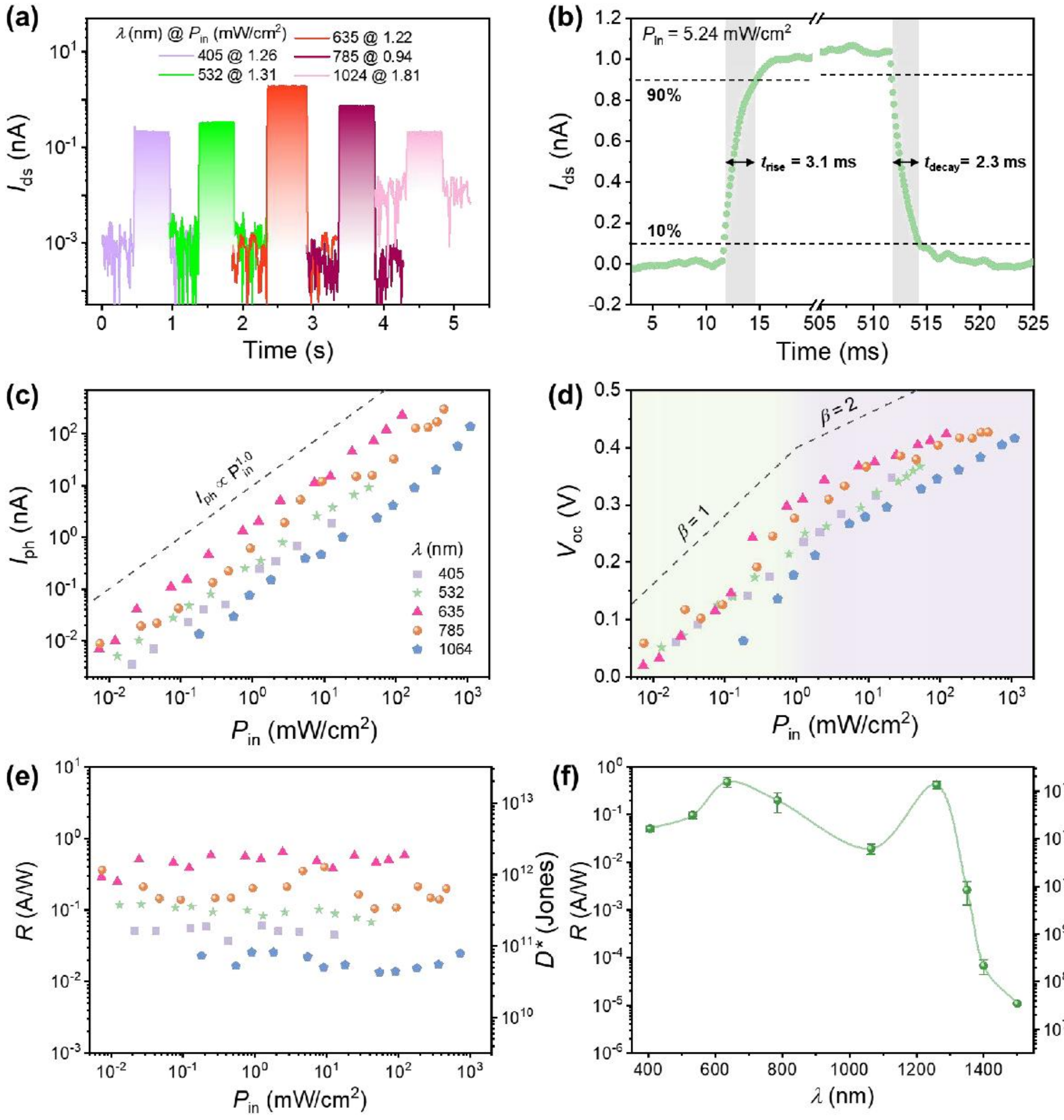

Fig. 4. (a) Time dependence of photocurrent at different wavelengths. (b) Response time of the device under 532 nm illumination when $P_{\mathrm{in}} = 5.24$ mW/cm$^2$. (c) $I_{\mathrm{ph}}$, (d) $V_{\mathrm{oc}}$ and (e) $R$ (left axis) and $D^*$ (right axis) versus $P_{\mathrm{in}}$. (f) The estimated $R$ (left axis) and $D^*$ (right axis) with respect to laser wavelength, where the data at 1260, 1350, 1400, and 1500 nm are extracted from Fig. S5.

Tunable photoresponse can be achieved by varying the pulsed gate voltage. Fig. 5a illustrates that the photocurrent increases from −0.096 to 0.68 nA under 532 nm illumination (laser power density 2.09 mW/cm$^2$) as $V_{\text{cg-pulse}}$ increases from −60 to 60 V (in 10 V increments). In comparison, the photocurrent observed in the $MoS_2$-based SFG-PD is two orders of amplitude smaller even under strong illumination (Fig. S9), suggesting its unsuitability for SFG-PDs utilizing unipolar 2D semiconductors. To understand the mechanism of this gate-tunability, spatially resolved scanning photocurrents were conducted using a focused 532 nm laser (spot diameter ~3 μm) at different $V_{\text{cg-pulse}}$. Fig. 5b-5c displays the photocurrent mapping of a typical device at $V_{\text{cg-pulse}} =- 60$, 10, and 60 V, respectively. When $V_{\text{cg-pulse}} =- 60$ V, the photocurrent mainly originates from two back-to-back Au/$MoTe_2$ Schottky junctions, confirming the configuration of $n^+$-n homojunction and Schottky junctions when $V_{\text{cg-pulse}} < 0$. Nevertheless, the contribution from the $n^+$-n homojunction is so weak that no discernible photocurrent is observed there. The corresponding band diagram is illustrated in Fig. 5d (left panel), where photogenerated electron-hole pairs are separated by the built-in electric field of the $n^+$-n homojunction and Au/$MoTe_2$ Schottky junctions, resulting in photocurrents of $I_1 > 0$, $I_2 < 0$, and $I_3 < 0$. The net photocurrent $I_{\text{ph}} = I_1 + I_2 + I_3$ is negative at $|I_{1/3}| \gg |I_2|$ and $|I_1| < |I_3|$. These two inequalities can be attributed, respectively, to the weak $n^+$-n homojunction and the strong tunnel effect at the left Schottky junction.[51] Theoretically, the contribution from the $n^+$-n homojunction can also be distinguished by reducing the two Schottky barriers, as evidenced by Fig. S10, which clearly shows negative photocurrent around the $n^+$-n homojunction at $V_{\text{cg-pulse}} =- 60$ V when using few-layer Gr as electrodes. For $V_{\text{cg-pulse}} = 10$ and 60 V, the photocurrent appears at the $MoTe_2$ homojunction, reaffirming the configuration of a p-n homojunction when $V_{\text{cg-pulse}} > 0$. Combined with the corresponding band diagram shown in Fig. 5d (right panel), the photocurrent of $I_1 < 0$, $I_2 > 0$, and $I_3 < 0$ can be inferred, leading to a net photocurrent of $I_{\text{ph}} > 0$ because $|I_{1/3}| \ll |I_2|$. The net photocurrent increases with increasing $|V_{\text{cg-pulse}}|$, attributed to a thinned Schottky barrier for $V_{\text{cg-pulse}} < 0$ and the enhanced built-in electric field in the p-n homojunction for $V_{\text{cg-pulse}} > 0$.

To assess the cycling and retention stability of the SFG-PD (Fig. S11a), we conducted transient photocurrent measurements under switchable laser illumination (1 Hz) for 5000 cycles, as illustrated in Fig. S11b-S11c. Clear on-off behavior was observed for both negative and positive gate voltage pulses. However, after thousands of cycles, the photocurrent for $V_{\text{cg-pulse}} = 60$ V appears to be more stable than that for $V_{\text{cg-pulse}} =- 60$ V, indicating that the electrons accumulated in the SFG are better confined than the holes. The extracted photocurrent data are presented in Fig. S11d, where the attenuation for $V_{\text{cg-pulse}} = 60$ V is noticeably milder than that for $V_{\text{cg-pulse}} =- 60$ V. Linear fitting in logarithmic coordinates revealed that the two photocurrents have reduced to 67.07% and 22.13%, respectively, after 10 years of measurement.

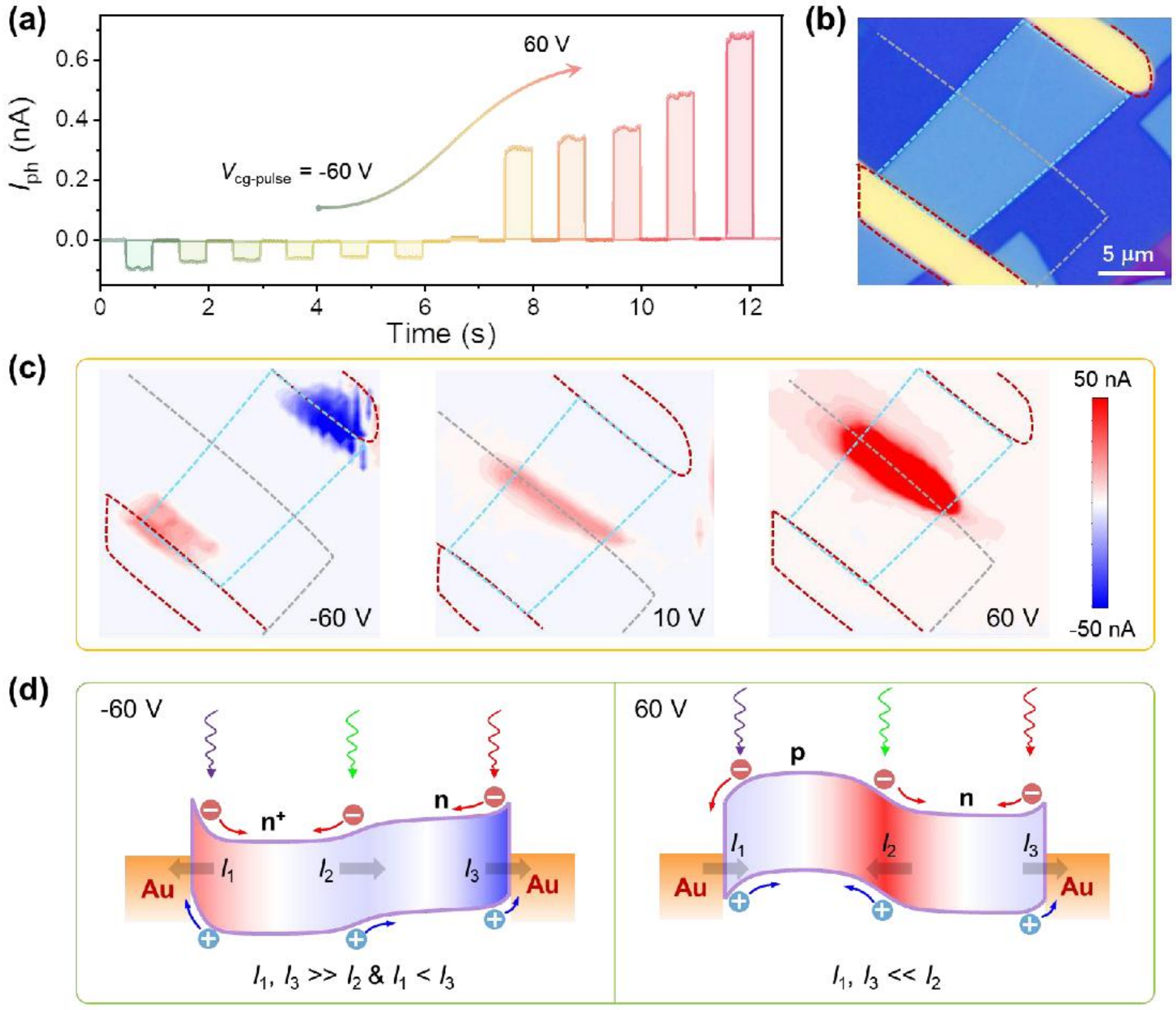

Fig. 5. (a) Time dependence of $I_{\text{ph}}$ at different $V_{\text{cg-pulse}}$ for $P_{\text{in}} = 2.09$ mW/cm$^2$. (b) Optical image of the $MoTe_2$/hBN/Gr SFG-PD with materials outlined in different colors. $MoTe_2$, Gr, and electrodes are outlined by cyan, gray, and crimson dashed lines, respectively. (c) Scanning photocurrent images at $V_{\text{cg-pulse}} =- 60$, 10, and 60 V, respectively, where the laser power is about 25 μW. (d) Schematic illustration of the carrier behavior and photocurrent production under illumination in the device at $V_{\text{cg-pulse}} =- 60$ and 60 V, respectively.

In summary, we successfully fabricated the nonvolatile $MoTe_2$/hBN/Gr SFG-PD with Au electrodes and conducted an in-depth investigation of its photodetection capabilities across a broadband spectrum. By leveraging the charge trapping of Gr SFG and the ambipolar characteristics of $MoTe_2$, the conducting channel can be finely tuned from $n^+$-n to p-n homojunction by increasing the pulsed gate voltage from negative to positive values. The device exhibits distinct rectification behavior in both configurations, with the rectification ratio for p-n being one or two orders of magnitude higher than that for $n^+$-n. This reconfiguration of the conducting channel allows us to modulate the photovoltaic current of the SFG-PD from negative to positive values. The positive photocurrent is primarily produced at the p-n homojunction, while the negative photocurrent stems from two back-to-back Au/$MoTe_2$ Schottky junctions (rather than the $n^+$-n homojunction). In the p-n configuration, the device demonstrates impressive photodetection performance with $V_{oc}$~0.4 V and $I_{sc}$~5.48 nA under 532 nm illumination at $P_{in} = 12.68$ mW/cm$^2$. Notably, this device delivers linear, rapid (~3 ms), and broadband (from 405 to 1500 nm) photodetection, with typical performances of $R$~0.5 A/W and $D^*$~$1.6 \times 10^{12}$ Jones at 635 nm. The performances of our device are favorably comparable to those of recently reported self-powered PDs of different types[52-61] (Fig. S12 and Table S1). Additionally, the pulsed gating demonstrates exceptional retention, with the photocurrent experiencing only a modest reduction to 67.07% after 10 years of operation. With these remarkable photodetection features, combined with its gate-tunability, the SFG-PD holds great promise for diverse applications in programmable optoelectronics[1, 2] and neuromorphic devices[3-5].

**Author Contributions**

F.L. and X.Luo conceived the concept and experiments and prepared the paper. F.L. fabricated and characterized the devices. X.Lin and Y.Y. assisted the experiments on device fabrication and measurements. X.G., Y.C., and X.Luo supervised and directed this project. All authors examined and commented on the paper.

**Notes**

The authors declare no competing financial interest.

**Acknowledgments**

This work was supported by the National Natural Science Foundation of China (Grant Nos. 61905198 and 62274087), the National Postdoctoral Program for Innovative Talents (No. BX20190283), and the Fundamental Research Funds for the Central Universities. Authors also thank the Analytical & Testing Center of NPU for the assistance in device fabrication.

***Supporting information***

## Experimental section

***Device Fabrication*:** All nanosheets of $MoTe_2$, hBN, and Gr were obtained with the mechanical exfoliation method from commercially available bulk crystals (HQ Graphene Company). The devices were fabricated by the dry transfer technique with the assistance of polydimethylsiloxane (PDMS) stamp. First, the Gr nanosheet was transferred on a $SiO_2$/Si substrate (highly n-doped Si with resistivity < 0.01 Ω cm, the $SiO_2$ thickness ~285 nm). Then, the hBN and $MoTe_2$ nanosheets were aligned and transferred in sequence on the Gr with the same method. The bottom Gr was covered by only part of the top $MoTe_2$ channel, forming a semi-floating gate configuration. Finally, Au films (thickness ~50 nm) were transferred on the $MoTe_2$ to form the electrodes. The as-prepared device was annealed at 200 °C in a vacuum tube furnace (BTF-1200C-S, BEQ) for 1 h to remove the resisted residues and to improve electrical contact.

***Characterizations*:** Raman spectra of $MoTe_2$, hBN, and Gr were measured by a confocal Raman microscope (Alpha300R, WITec) in ambient condition. The samples were excited by a 532 nm laser (spot size ~400 nm, laser power ~1 mW, resolution ~0.02 $cm^{-1}$). AFM measurements were conducted using an atomic force microscope (Dimension FastScan, Bruker). The electrical/optoelectrical measurements of device under 405, 532, 635, 785, and 1064 nm laser illumination were performed at room temperature by a semiconductor parameter analyzer (FS380 Pro, Platform Design Automation) in a probe station with a vacuum degree of ~$5\times10^{-6}$ mbar. The optoelectrical measurements of device under 1260, 1350, 1400, and 1500 nm laser illumination were performed in ambient condition by a dual-channel source meter (B2912A, Keysight). The photocurrent mapping was measured in ambient condition by a home-built photocurrent scanning system with the diameter of the focused 532 nm laser spot ~3 μm.

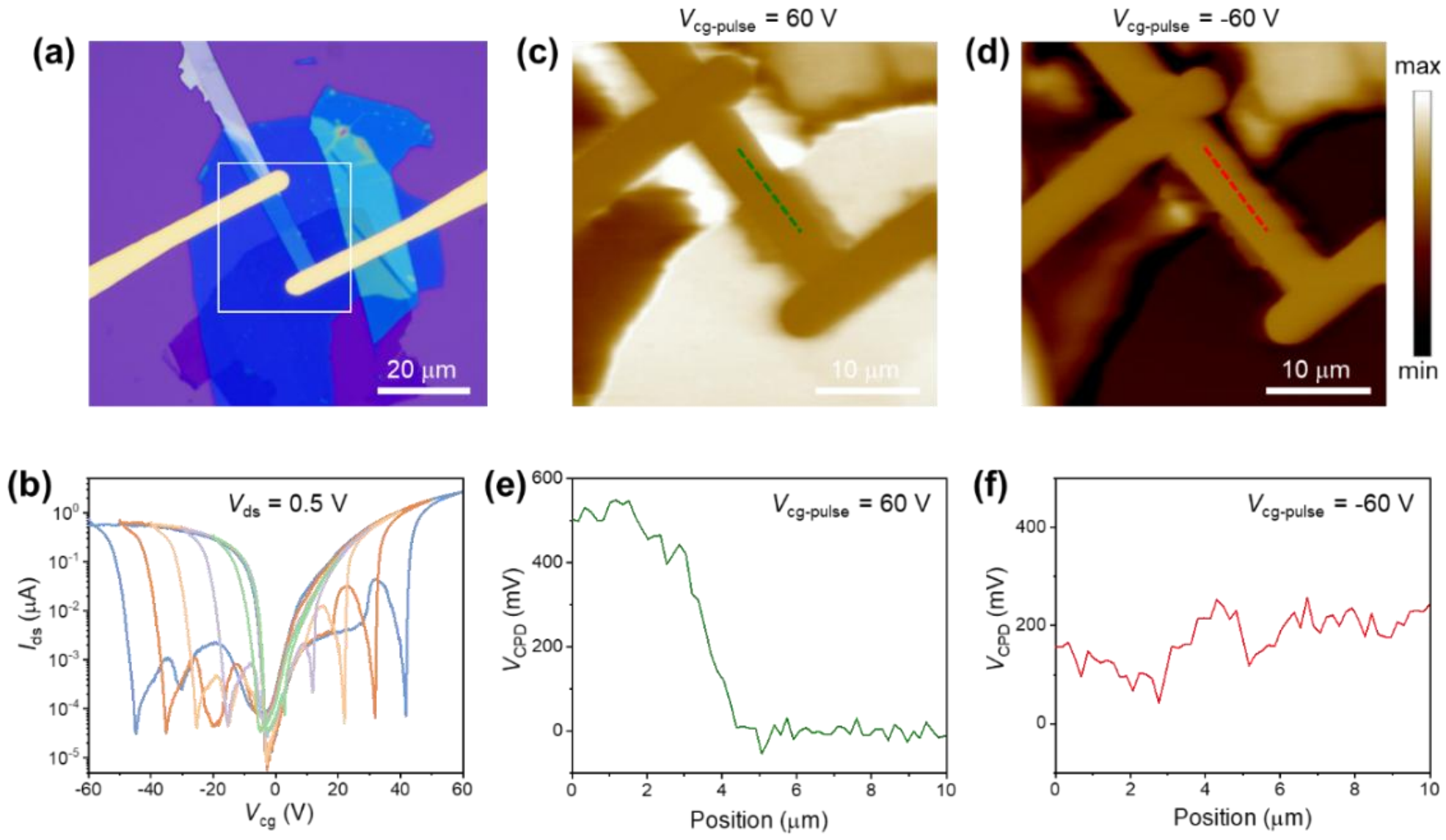

Fig. S1. (a) The optical image of the device; (b) Transfer curves as the $V_{cg}$ scanning in the range of ±20, ±30, ±40, ±50, and ±60 V, respectively; (c) KPFM of the white box in (a) when the pulse voltage of $V_{cg}$ is 60 V; (d) KPFM of the white box in (a) when the pulse voltage of $V_{cg}$ is -60 V; (e) and (f) are the contact potential difference profile along the dashed lines in (c) and (d), respectively, and left sides are free of SFG. The contact potential difference is $V_{CPD} = V_{tip} - V_{sample} = (W_{sample} - W_{tip})/q$, where $V$ and $W$ indicate voltage and work function, respectively.

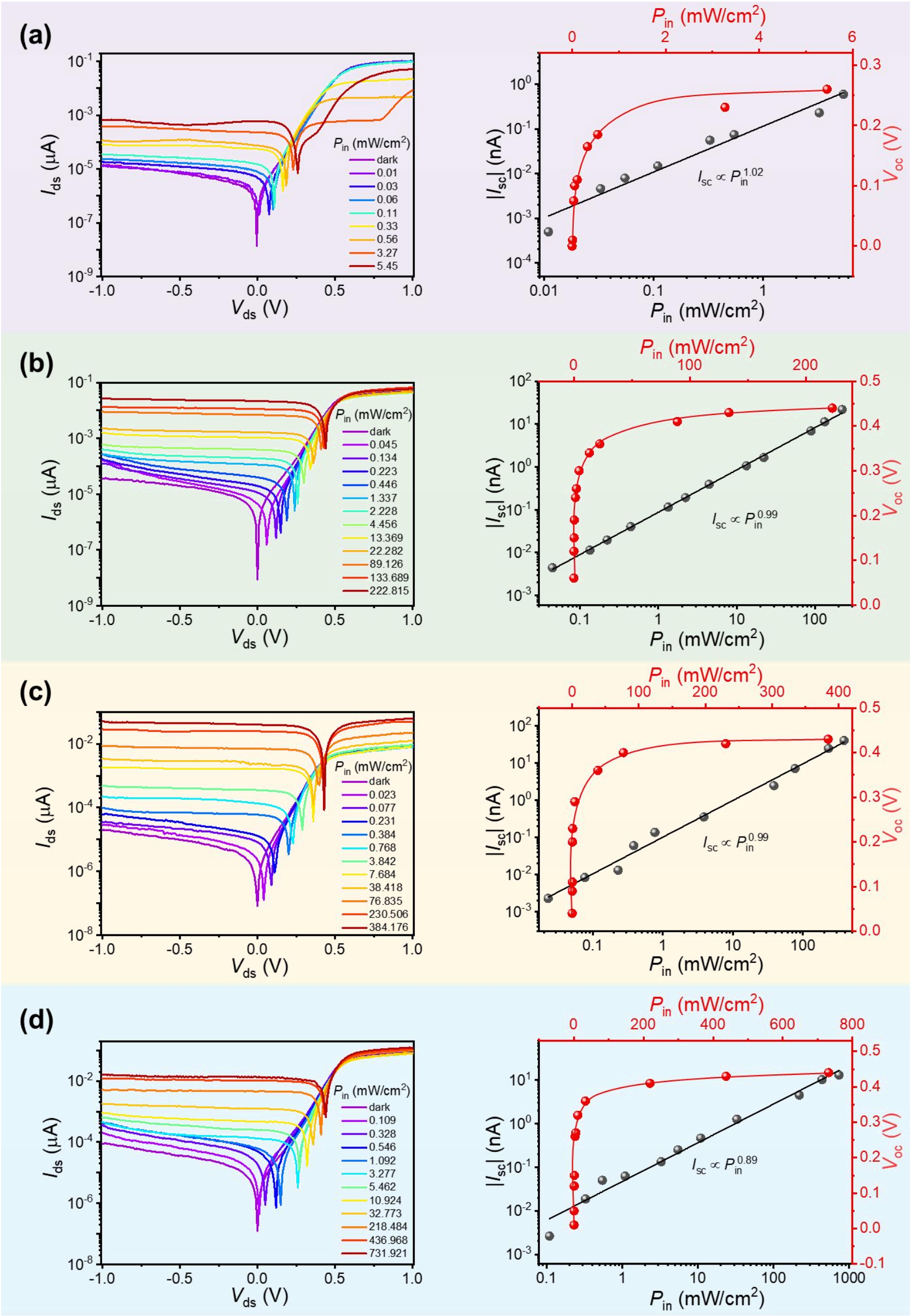

Fig. S2. Semi-log plot of output curves ($I_{\mathrm{ds}} - V_{\mathrm{ds}}$) of the SFG-PD under illumination with different $P_{\mathrm{in}}$ (left panel) when $V_{\mathrm{cg-pulse}} = 60$ V, and the extracted $I_{\mathrm{sc}}$ and $V_{\mathrm{oc}}$ as a function of $P_{\mathrm{in}}$ on logarithmic (left axis) and linear (right axis) scales (right panel). The laser wavelengths are (a) 405 nm, (b) 635 nm, (c) 785 nm, and (d) 1064 nm, respectively.

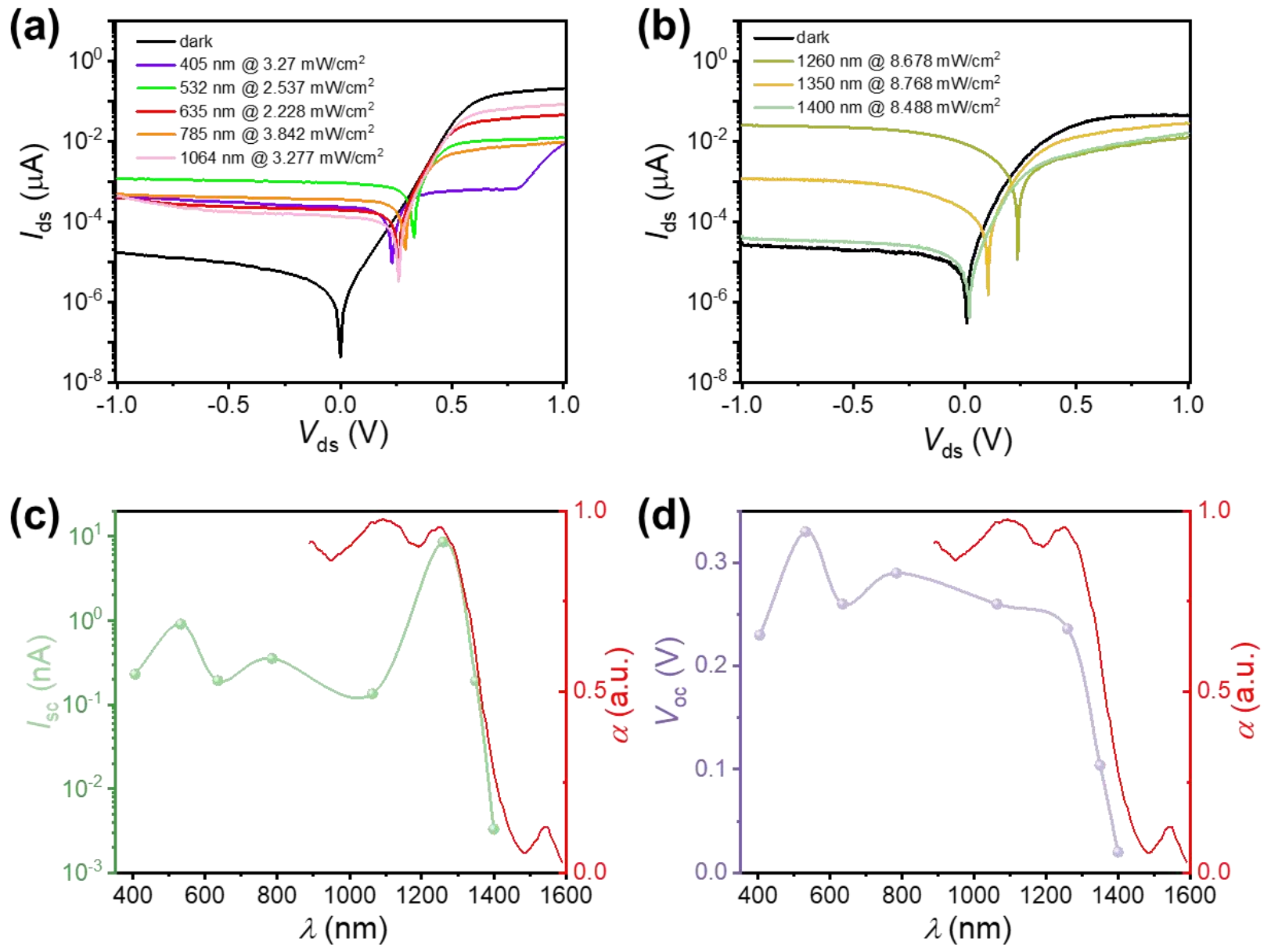

Fig. S3. Semi-log plot of output curves of the SFG-PD under illumination with different lasers when $V_{\mathrm{cg-pulse}} =$ 60 V: (a) measured in the vacuum chamber; (b) measured at ambient condition, where the data for 1500 nm is not shown here because of the small variation from the dark data. The extracted (c) short-circuit current and (d) open-circuit voltage with respect to the laser wavelength. The red lines in (c) and (d) are the absorption of few-layer $MoTe_2$ in previous report.[1]

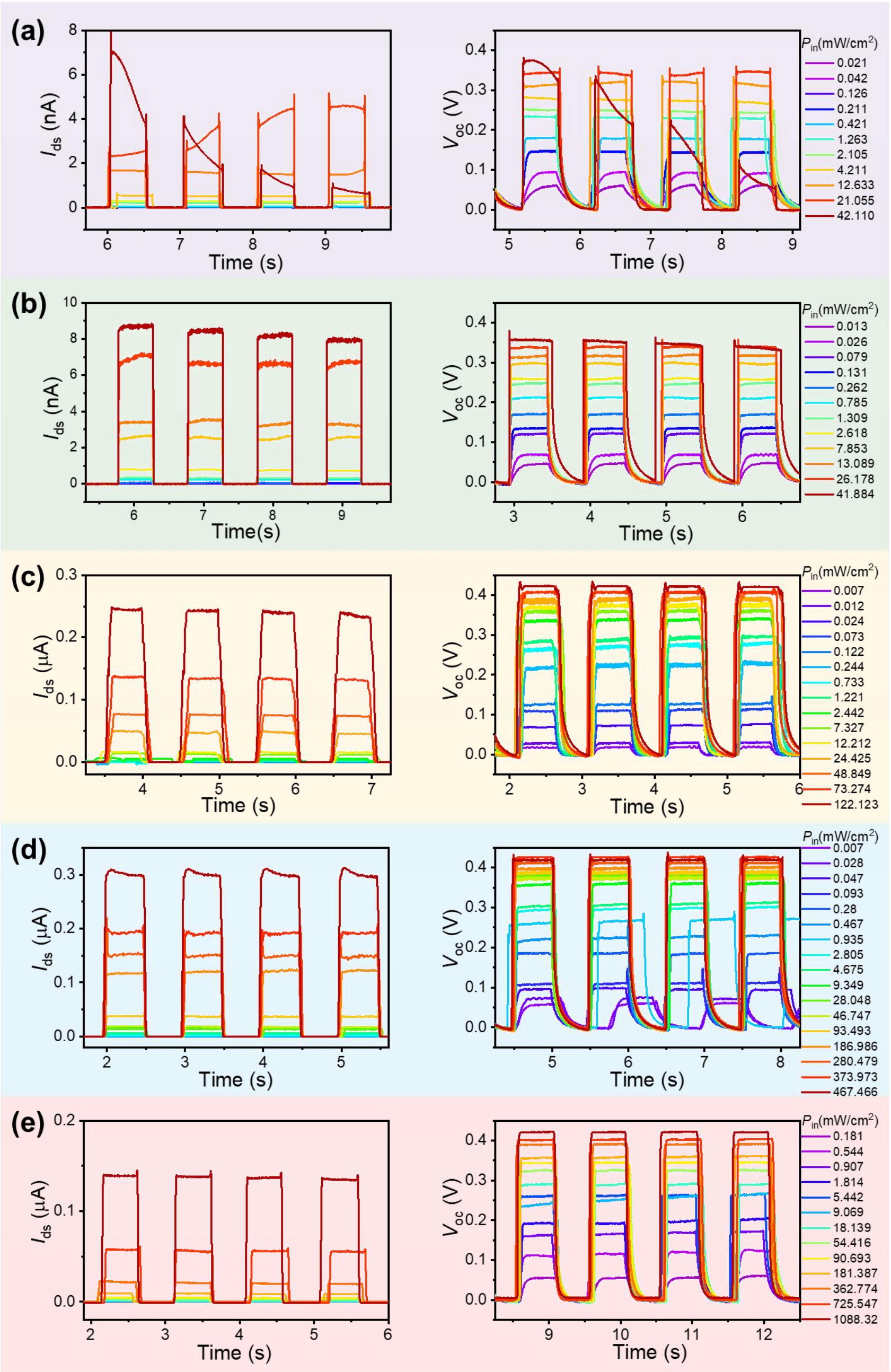

Fig. S4. Transient response of photocurrent (left panel) and $V_{\text{oc}}$ (right panel) of the SFG-PD under illumination with different lasers when $V_{\text{cg-pulse}} = 60$ V. (a) 405 nm; (b) 532 nm; (c) 635 nm; (d) 785 nm; (e) 1064 nm.

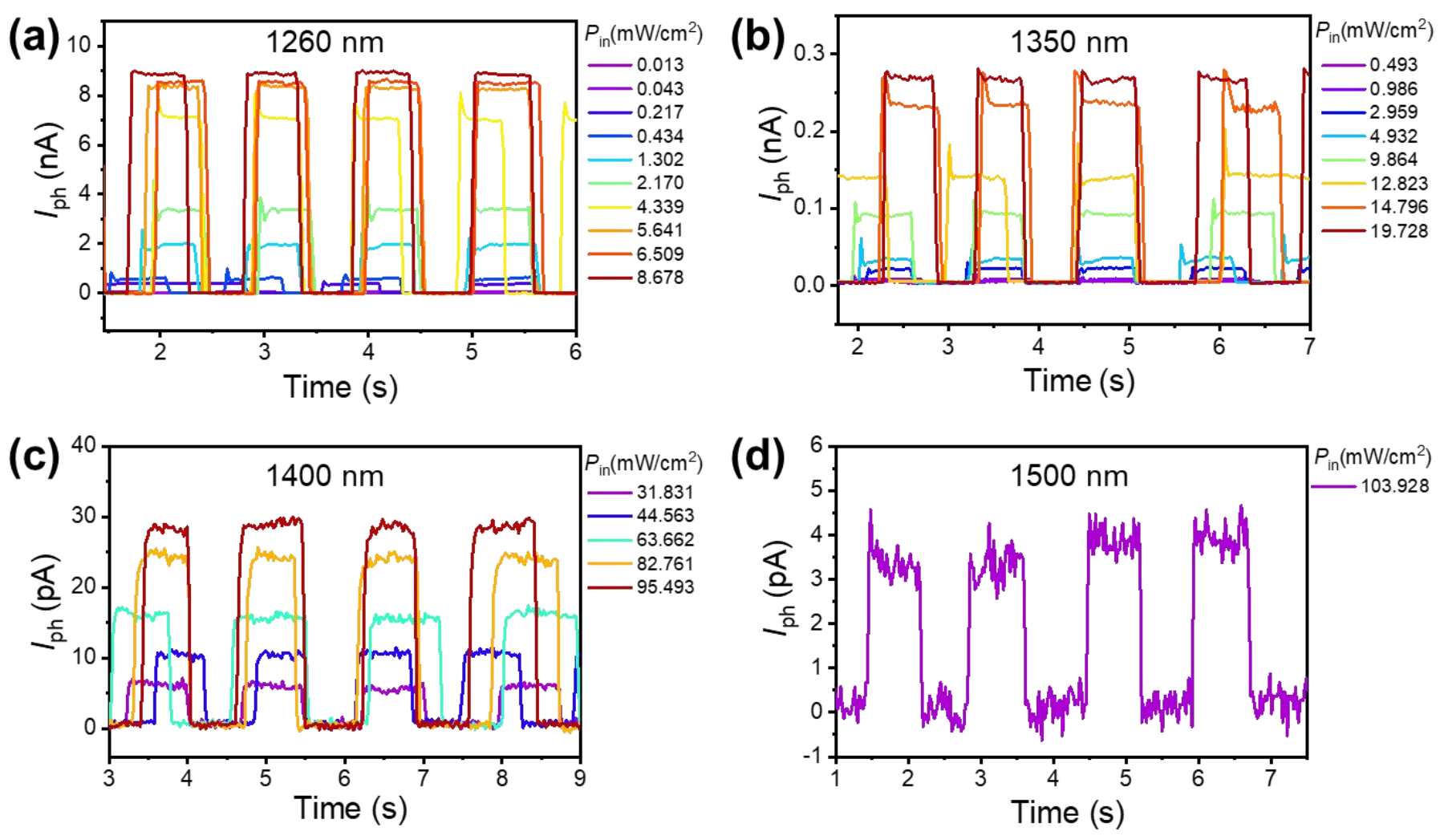

Fig. S5. Transient photocurrent of the SFG-PD measured in the air under different illumination when $V_{\mathrm{cg-pulse}} =$ 60 V. (a) 1260 nm; (b) 1350 nm; (c) 1400 nm; (d) 1500 nm.

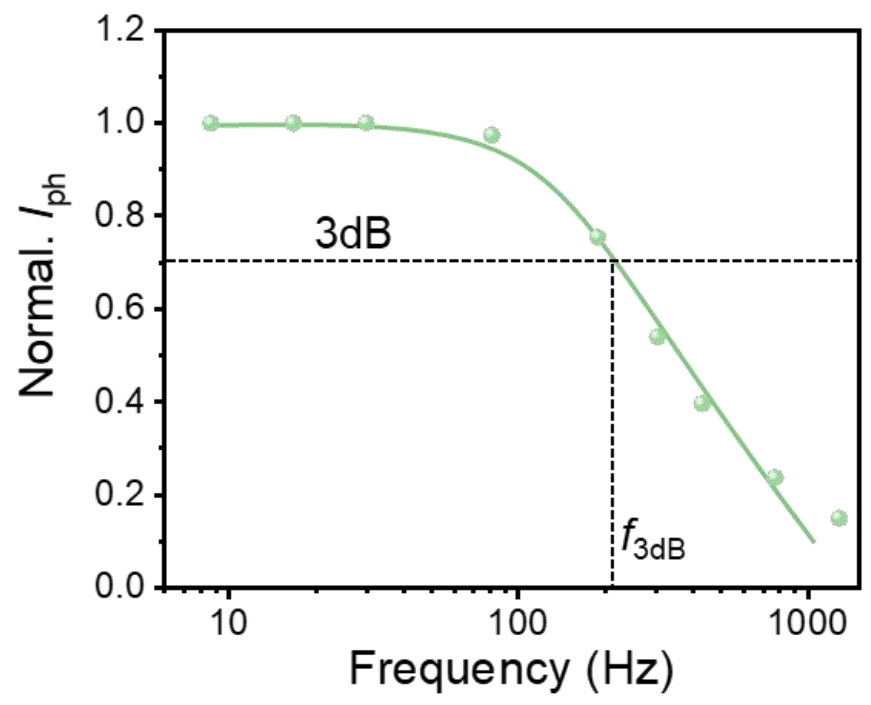

Fig. S6. The measured 3dB bandwidth of the SFG-PD under 532 nm illumination when $V_{\mathrm{cg-pulse}} = 60$ V.

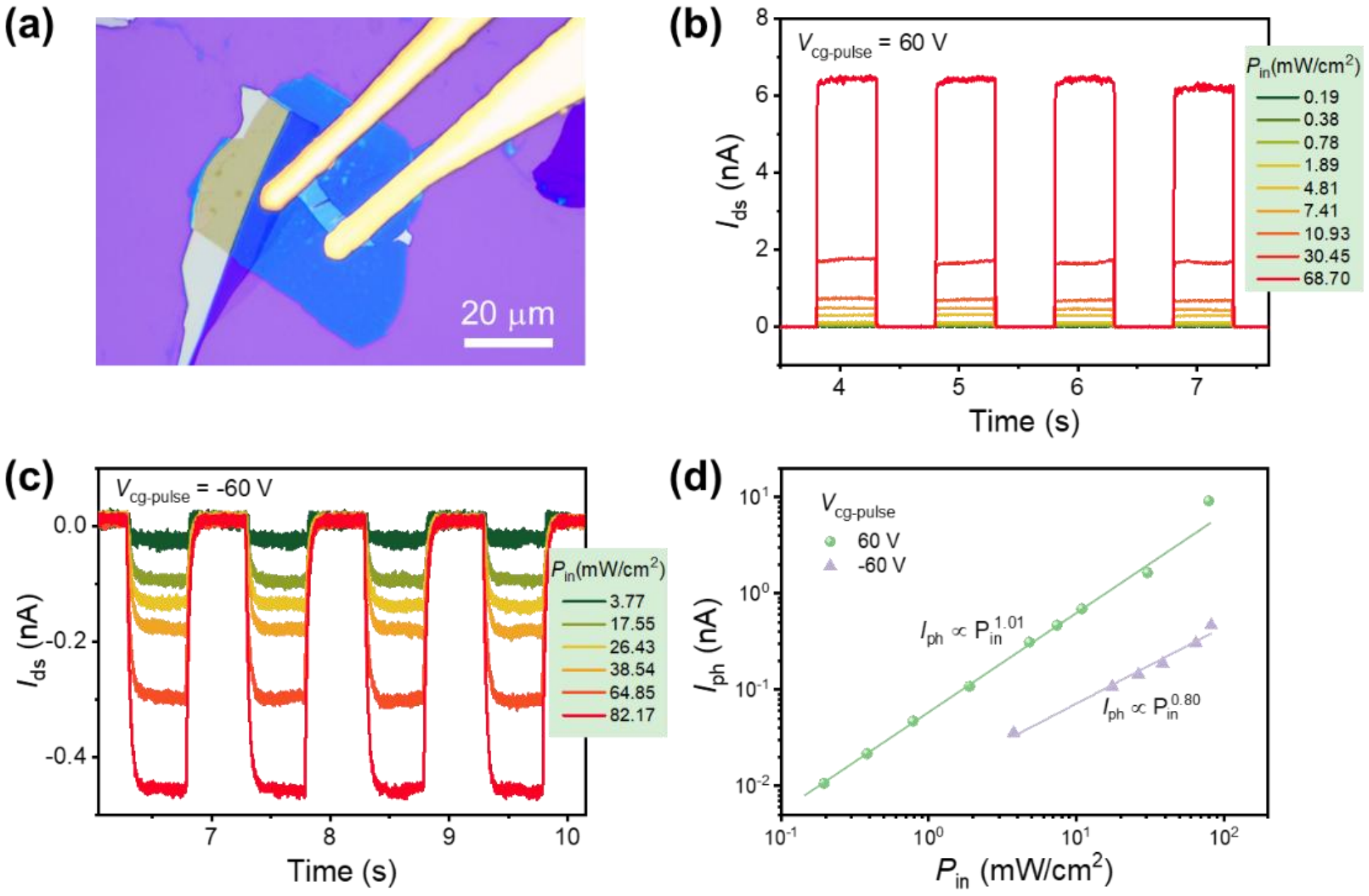

Fig. S7. (a) The optical image of an SFG-PD. Transient response of photocurrent of a under 532 nm illumination when (b) $V_{\mathrm{cg-pulse}} = 60$ V and (c) $V_{\mathrm{cg-pulse}} =- 60$ V. (d) The extracted photocurrent with respect to the laser power density, where the solid lines are the fitted results.

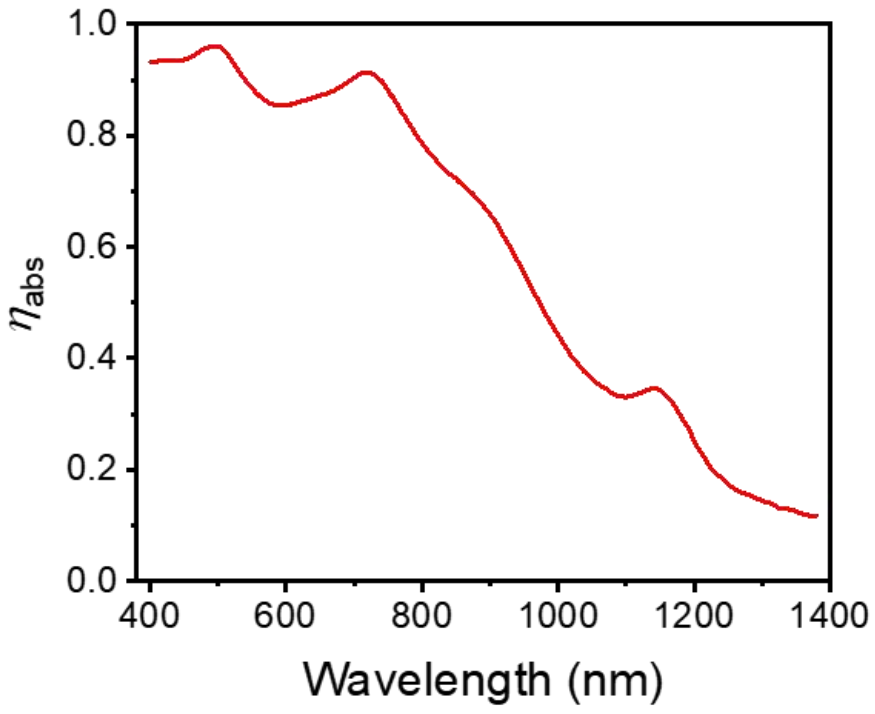

Fig. S8. The calculated absorbance of 6.3 nm $MoTe_2$ based on Lambert–Beer law of $\eta_{abs} = 1 - e^{-\alpha d}$, where $d$ is the $MoTe_2$ thickness and the absorption coefficient $\alpha$ is calculated with the experimental data of trilayer $MoTe_2$.[2]

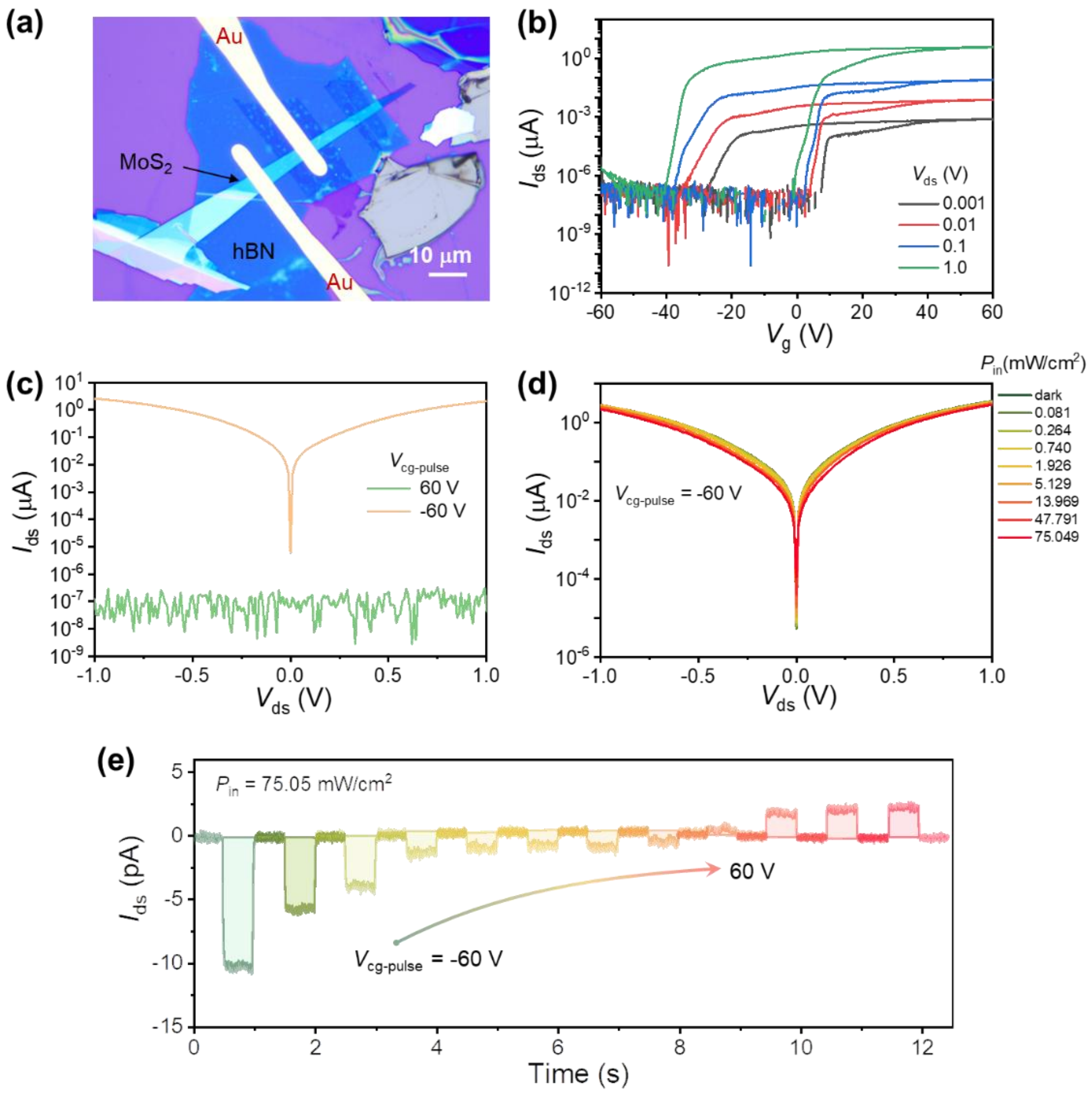

Fig. S9. (a) Optical image of a $MoS_2$/hBN/graphene SFG-PD. (b) Transfer curves with continuous gate voltage (loop scanning) at different $V_{ds}$. (c) Output curves under $V_{cg-pulse} = 60$ and $-60$ V, with the pulse duration of 1 s. (d) No discernible open-circuit voltage observed in the output curves under 532 nm illumination when $V_{cg-pulse} = -60$ V. (e) Transient photocurrent as $V_{cg-pulse}$ ranges from $-60$ to 60 V under $V_{ds} = 0$ V and strong illumination ($P_{in} = 75.05$ mW/cm$^2$).

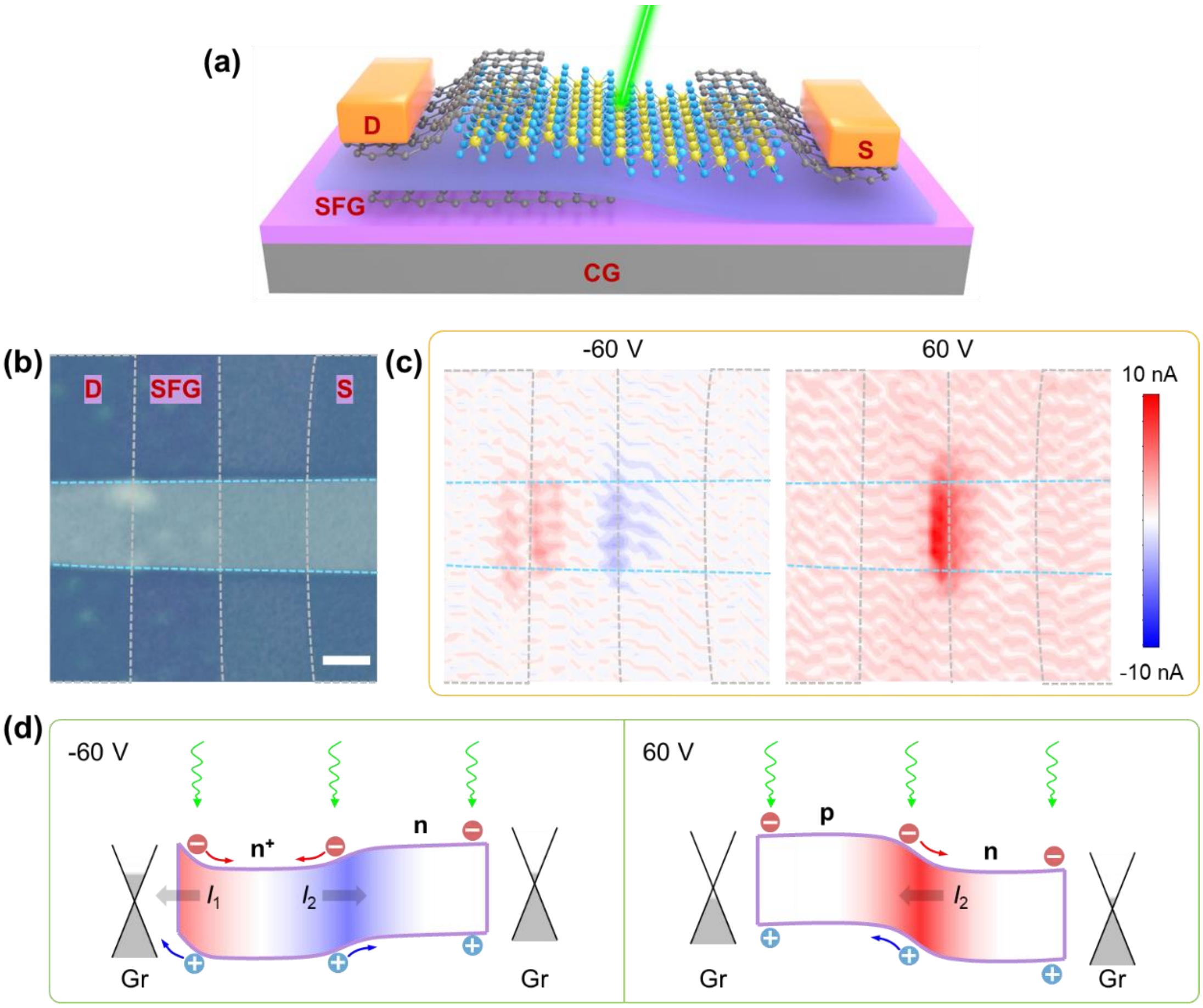

Fig. S10. (a) Schematic of the SFG-PD with Graphene electrodes. (b) Optical image of the typical device, where $MoTe_2$ and graphene are outlined by cyan and gray dashed lines, respectively. Scale bar: 2 μm. (c) Scanning photocurrent images at $V_{\mathrm{cg-pulse}} = -60$ and 60 V, respectively, under activated by the focused 532 nm laser, where the laser power is about 43 μW. (d) Schematic illustration of the carrier behavior and photocurrent production under illumination of the device at $V_{\mathrm{cg-pulse}} = -60$ and 60 V, respectively.

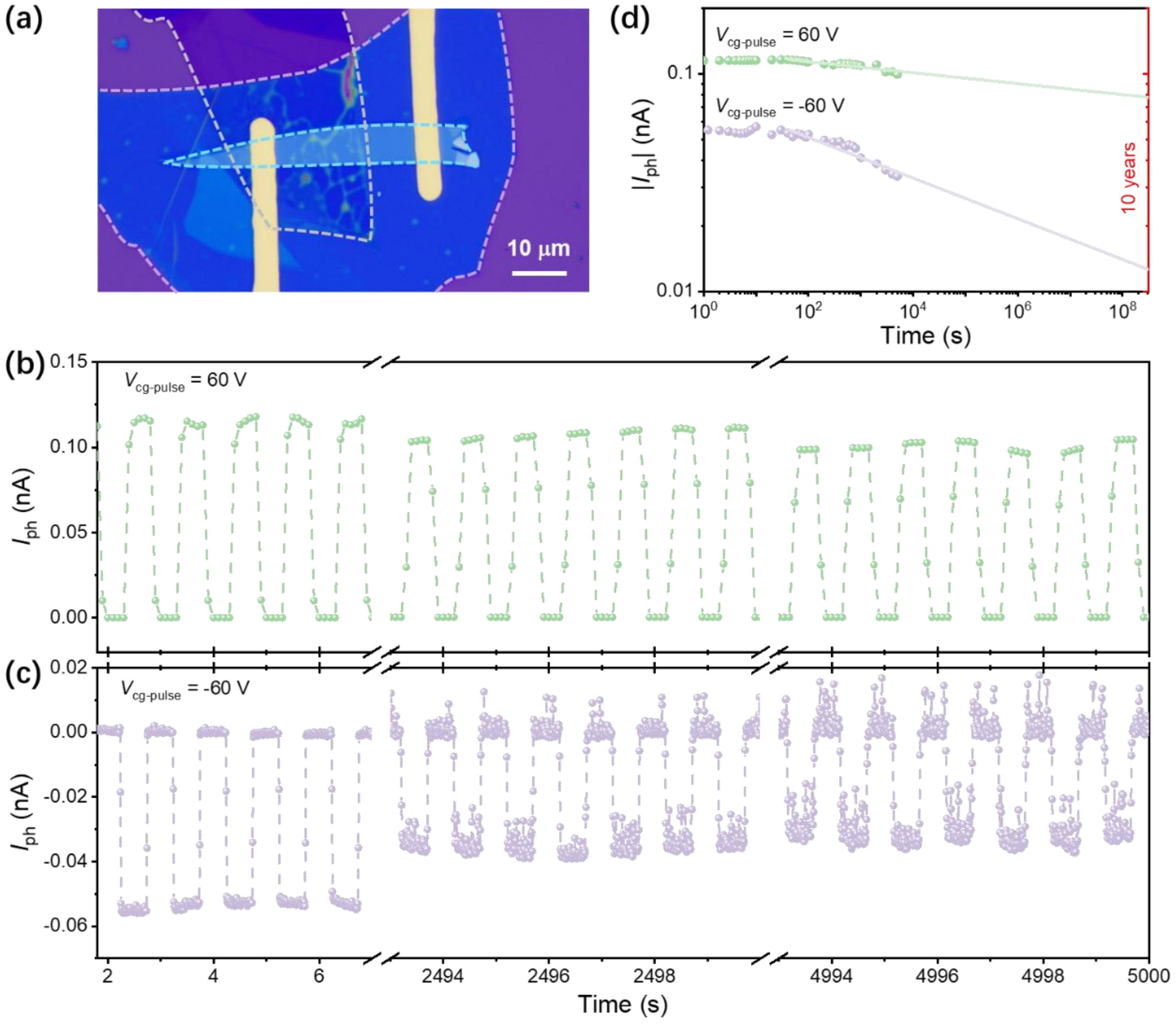

Fig. S11. The retention time of the SFG-PD under 532 nm illumination. (a) Optical image of the measured device. (b-c) Transient response of photocurrent lasting 5000 s when $V_{cg-pulse} = \pm 60$ V and $P_{in} = 1.68$ mW/cm$^2$. (d) The extracted photocurrent with respect to time, and the fitted value after 10 years are lowered to 67.07% and 22.13% for $V_{cg-pulse} = 60$ and $-60$ V, respectively.

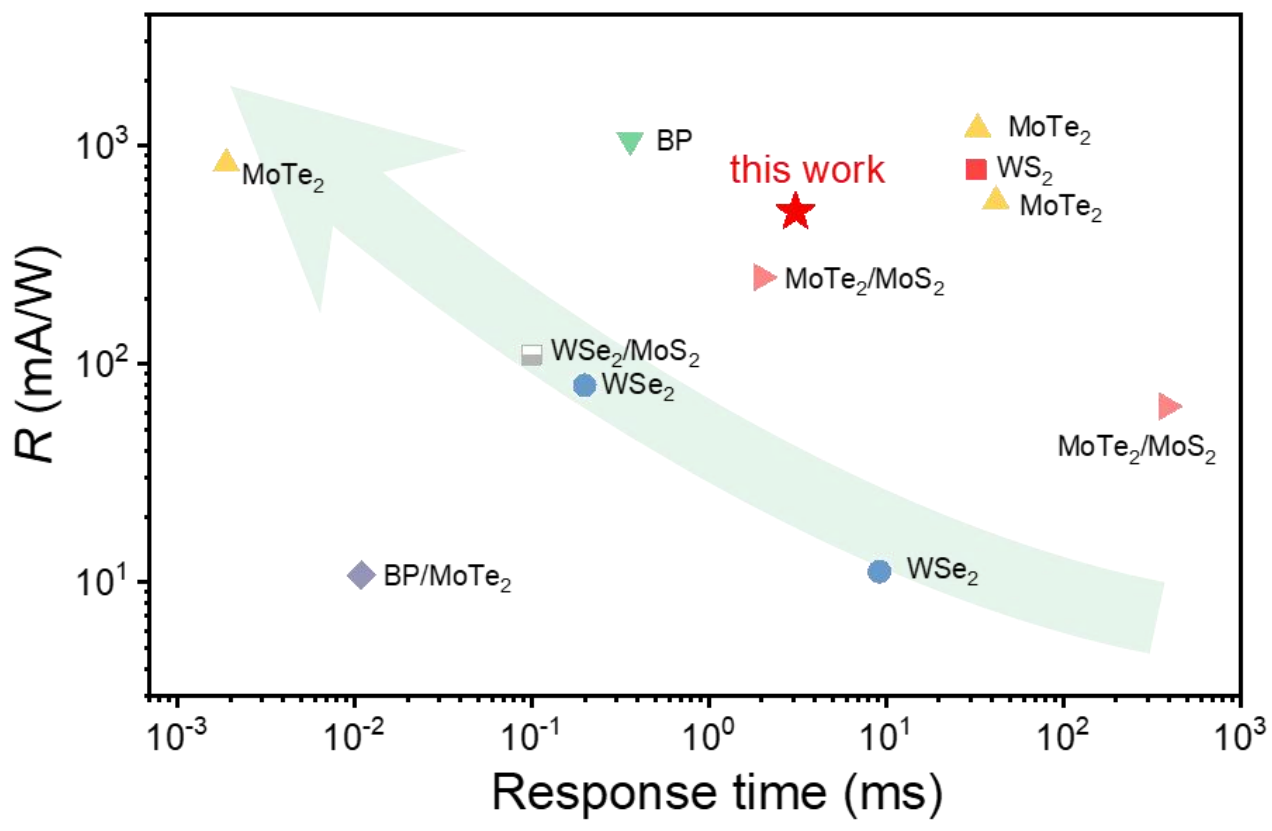

Fig. S12. Responsivity versus response time for several other self-powered 2D photodetectors based on the data presented in Table S1.

Table S1. Performance parameters of recently reported self-powered photodetectors based on 2D materials.

| Device | Thickness (nm) | $\lambda$ (nm) | $V_g$ (V) | $R$ (mA/W) | $D^*$ (Jones) | $t$ (ms) | Ref. |
|---|---|---|---|---|---|---|---|
| $WSe_2$ | 90 | 520 | -40 | 80 | $10^{11}$ | 0.2 | 3 |
| $WS_2$ | 81.6/104 | 405 | 0 | 777 | $10^{11}$ | 32.3 | 4 |
| $WSe_2$ | 6/65 | 532 | 0 | 11.2 | $10^{10}$ | 9.22 | 5 |
| $MoTe_2$ | 19.93/7.86 | 635/1064 | 40 | 1200/560 | $10^{12}/10^{11}$ | 32.96/41.96 | 6 |
| $MoTe_2$ | - | 635 | 25/-25 (pulse) | 830 | - | 0.0019 | 7 |
| BP | 4.8 | 1450 | -3/3 (split) | 1060 | $10^{11}$ | 0.362 | 8 |
| BP/$MoTe_2$ | 15/23 | 532 | 0 | 10.8 | - | 0.011 | 9 |
| $WSe_2/MoS_2$ | 25/18 | 532 | 0 | 110 | - | <0.1 | 10 |
| $MoTe_2/MoS_2$ | 10/13 | 600 | 0 | 250 | - | 2 | 11 |
| $MoTe_2/MoS_2$ | 1.5/3.8 | 473 | 80 | 64 | $10^{10}$ | 385 | 12 |
| $MoTe_2$ | 6.3 | 635 | 60 (pulse) | 500 | $10^{12}$ | 3.1 | this work |